\documentclass[10pt,letterpaper]{article}
\usepackage{opex3}

\begin{document}

\title{The Correlation Confocal Microscope}

\vskip4pc


\title{The Correlation Confocal Microscope}

\author{D.S. Simon$^{1.}$ and A.V. Sergienko$^{1,2}$}

\address{$^1$ Dept. of Electrical and Computer Engineering, Boston
University, 8 Saint Mary's St., Boston, MA 02215.

$^2$ Dept. of Physics, Boston University, 590 Commonwealth Ave.,
Boston, MA 02215}

\email{simond@bu.edu} 



\begin{abstract} A new type of confocal microscope is described which makes use of
intensity correlations between spatially correlated beams of
light. It is shown that this apparatus leads to significantly
improved transverse resolution.\end{abstract}



\section{Introduction}

The development of the confocal microscope \cite{1,2}
led to a revolution in microscopy. The insertion of source and
detection pinholes leads to improved resolution and contrast and
the ability to image thin optical sections of a sample in a
noninvasive manner. This has made the confocal microscope
ubiquitous in modern biomedical optics research, where it is vital
not only for imaging, but also for dynamic light scattering
(\cite{3,4,5}), fluorescent correlation
spectroscopy (\cite{6,7}), and other types of experiments.
Because all of these experiments rely on achieving the smallest
confocal volume (the overlap of the images in the sample of the
source and detector pinholes), a great deal of effort has gone
into improving the resolution of confocal microscopes in order to
minimize this volume. One common approach, which exploits the idea
of correlated excitation, is two photon microscopy
\cite{8,9}, in which a pair of photons must be absorbed by
a fluorescent molecule simultaneously; since this only happens
with appreciable probability where the photon density is very
high, only the central part of the confocal volume contributes,
leading to a reduced {\it effective} confocal volume.

Separately, Hanbury Brown and Twiss
[10-12] in the 1950's showed that the
use of intensity correlations on the detection side could improve
the resolution of radio and optical measurements in astronomy. In
recent years, similar intensity correlation methods have been used
extensively in quantum optics, leading to a number of developments
including quantum ellipsometry
[13-15], correlated two-photon or ghost
imaging \cite{16}, and aberration and dispersion cancellation
[17-22].
All of these effects were first discovered using quantum entangled
photon pairs produced via parametric downconversion, but some have
since been reproduced using beams of light with classical spatial
correlations; see for example
[23-30].

In this paper we investigate the question of whether
the advantages of intensity correlation
methods at detection and of confocal microscopy can be combined.
We will
show that use of transverse (lateral) spatial correlations and of
coincidence detection can significantly improve the resolution of
a confocal microscope. Some care needs to be taken in how this is
done since following the naive method of simply sending pairs of
spatially correlated photons into a standard confocal microscope
will not work; the pinhole destroys all spatial correlations. So
instead, we send in uncorrelated photons and use a form of
postselection to enforce correlations among the photons we choose
to detect.

It is to be noted that the correlation method described here
shares in a sense a common underlying philosophy with two-photon
microscopy, since the two-photon microscope also uses spatial
correlations, but at the {\it excitation stage}: uncorrelated photons
are inserted into the microscope at the source, but the
requirement that the two photons interact with the same
fluorescent molecule effectively enforces spatial correlations
among the photon pairs that contribute to the detected signal. We
do something similar here, but use a different method in order to
enforce the correlations at the {\it detection stage}. This will reduce
the need for high intensities, thus allowing the use of less
powerful lasers, as well as reducing possible damage to the
sample.

It should also be pointed out that the idea of using intensity
correlations with entangled photon pairs has been been applied
before to obtain subwavelength microscopic resolution
\cite{31}, though not in conjunction with confocal
microscopy. This previous work, known as quantum microscopy,
required entangled states, in contrast to the method here which
will work with a completely classical light source.

The outline of the paper is as follows. In section 2, confocal
microscopy is briefly reviewed.  Section 3 discuss the problem
of combining confocality with correlation. In section 4 we
introduce the setup for a generalized version of the correlation
confocal microscope, and show that it does indeed lead to
significant improvement in resolution over the standard
(uncorrelated) confocal microscope. In order to make the
principles of operation clearer, we will initially consider in
section 4 the unrealistic case of a generalized microscope that
requires two identical copies of the object being viewed; in
section 5, we then show how to reduce the apparatus to the
realistic case of a single object. Section 6 looks at the axial
resolution, with conclusions following in section
7.

\section{The Standard Confocal Microscope.}

The basic setup of a standard confocal microscope is shown in
figure 1. (For a more detailed review see \cite{32}). The two
lenses are identical. In real setups they are in fact usually
the {\it same} lens, with reflection rather than transmission
occurring at the sample. (In this paper we will for simplicity
always draw the transmission case, but most of the considerations
will apply equally to the reflection case.) This lens serves as the objective; it has
focal length $f$ and radius $a$, and serves to focus the light
going in and out of the sample. The sample is represented at point
${\mathbf y}$ by a function $t({\mathbf y})$; depending on the
setup, $t({\mathbf y})$ will represent either the transmittance or
reflectance of the sample. At the first lens, the distances are
chosen so that the imaging condition \begin{equation} {1\over
{z_1}}+{1\over {z_2}}={1\over f}\end{equation} is satisfied; as a
result, light entering the microscope through the source pinhole
is focused to a small diffraction-limited disk (actually a
three-dimensional ellipsoid) centered at a point $P$ in the
sample. Any stray light not focused to this point is blocked by
the pinhole, thus providing the first improvement in contrast
between between $P$ and neighboring points. The distances at the
second lens also satisfy the imaging condition, so the second lens
performs the inverse of the operation carried out by the first
one, mapping the diffraction disk in the sample back to a point at
the detection plane. The pinhole in this plane blocks any light
not coming from the immediate vicinity of $P$, thus providing
further contrast. Together, the two pinholes serve to pass light
from a small in-focus region in the sample
and to block light from out-of-focus
regions. The in-focus point is then scanned over the sample.
The end result is a significant improvement in contrast
over the widefield microscope.

The double passage through the lens also leads to improved
resolution. To quantify the resolution improvement, we need to
look at the impulse response function $h({\mathbf y})$ and
transverse point-spread function (PSF) of the microscope. Let
$h_i(\xi ,y)$ ($i=1,2$) be the impulse response functions for the
first and second lenses individually (including the free space
propagation before and after the lens). Up to multiplication by
overall constants, these are of the form
\begin{equation} h_i({\mathbf \xi} ,{\mathbf y})= e^{{{ik}\over 2}\left( {{y^2}\over {z_2}}+{{\xi_2^2}\over {z_1}}\right)}\tilde p\left(
k\left( {{\mathbf y}\over {z_2}}+{{\mathbf \xi}\over {z_1}}\right)
\right) ,\end{equation} where $\tilde p({\mathbf q})$ is the
Fourier transform of the aperture function $p({\mathbf x^\prime})$
of the lens. ${\mathbf q}$ and $k$ respectively denote the
transverse and longitudinal momenta of the incoming photon. We
assume that $q<<k$. From now on, we will also assume a circular, abberation-free lens of radius $a$, in which case:
\begin{equation} \tilde p({\mathbf q})=2\pi a^2{{J_1\left(a q \right)}\over {(a q)}}, \label{ptil}\end{equation}
where $q$ is the magnitude of $ {\mathbf q}$ and $J_1$ is the
Bessel function of first order. Applying a pinhole at one end, we
may also define
\begin{equation}h_i({\mathbf y})=h_i(0 ,{\mathbf y}) =e^{{{iky^2}\over {2z_2}}}\tilde p \left( {{ky}\over {z_2}}\right)\label{stage1}\end{equation}

\begin{figure}
\centering
\includegraphics[width=7cm]{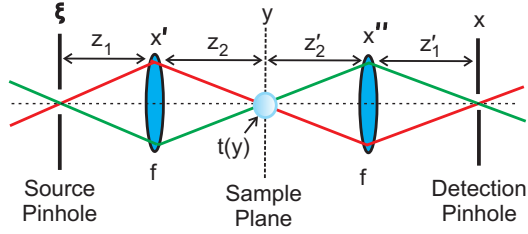}
\caption{\textit{(Color online) Schematic diagram for standard
confocal microscope.}}
\end{figure}

Imagine that a sample is being scanned by the microscope. The amplitude impulse
response for the microscope while focused at sample point ${\mathbf y}$ is
\begin{eqnarray}h({\mathbf y})&=&\int d^2y^\prime \; h_1({\mathbf \xi} ,{\mathbf y^\prime } )
t({\mathbf y^\prime }+{\mathbf y}) h_2({\mathbf  y^\prime} ,{\mathbf x} )\left|_{{\mathbf x}={\mathbf \xi}=0}\right. \\
&=& \int d^2y^\prime \; h_1({\mathbf y^\prime} ) t({\mathbf
y^\prime}+{\mathbf y}) h_2({\mathbf y^\prime} )  .
\end{eqnarray}

If we insert a sample which is nontransmitting except at a single
point, $t({\mathbf y})=\delta ({\mathbf y})$, the impulse response
becomes the coherent spread function
\begin{equation}CSF({\mathbf y})= h_1(-{\mathbf y})h_2(-{\mathbf
y}),\end{equation} with the corresponding point-spread function
given by
\begin{equation}PSF({\mathbf y})= \left| h_1(-{\mathbf y})h_2(-{\mathbf
y})\right|^2 .\end{equation} Setting the distances at the first
and second lenses to be equal for simplicity ($z_i^\prime =z_i$
for $i=1,2$), use of equation (\ref{ptil}) gives us
(\cite{32,33}):
\begin{equation} PSF({\mathbf y})\propto \left[ \tilde p\left( {{k{\mathbf y}}\over
{z_2}}\right)\right]^4 \propto \left( {{J_1\left( {{kay}\over
{z_2}}\right) }\over {\left( kay/z_2\right)}}\right)^4
,\end{equation} to be compared with the widefield microscope,
which has
\begin{equation}PSF_{wide}({\mathbf y})\propto\left[ \tilde p\left( {{k{\mathbf
y}}\over {z_2}}\right)\right]^2\propto\left( {{J_1\left( {{kay}\over
{z_2}}\right) }\over {\left(
kay/z_2\right)}}\right)^2.\end{equation} Since $\tilde p$ is a
sharply-peaked function, the higher power in the confocal result
leads to a further increase of sharpness and a resulting
improvement in resolution.


\section{Correlations versus Confocality.}

We now wish to introduce spatial correlations into the confocal
microscope. However, we immediately run into a problem: the source
pinhole is ideally a delta function in position space, which means
that its Fourier transform is a constant in momentum space. So,
regardless of the spectrum of transverse spatial momentum
${\mathbf q}$ entering the pinhole, the spectrum leaving the
pinhole is approximately flat; all transverse momentum
correlations are lost. Thus it seems that we must choose between
keeping either the correlations or the source pinhole, but not
both.

However, the problem may easily be avoided in several ways. One
category of solution involves removing the source pinhole and
preserving the confocality by other means. Notice that the only
purpose the source pinhole serves is to make sure that all of the
light entering the microscope is focused at the same point in the
object plane. But this can be achieved without a pinhole and therefore without
destroying spatial
correlations. This can be done in several ways:

(1) One method, often used in the standard
confocal microscope, is to have the beam hit the lens parallel to
the axis and focus to a point one focal length $f$ away (figure
2a). We can then introduce correlations by arranging for pairs of
narrow, well-localized beams of light to strike the lens at equal
distances from the axis.

(2) A second method is to use pairs of photons
produced by spontaneous parametric downconversion at correlated
angles, so that if they are traced backward they both seem to
emanate from the same point (figure 2b); that point would be then
be analogous to the pinhole. The crystal would have to be far enough
from the lens for the cross-section of the pump beam to appear to be pointlike.

(3) A third method is to have a
very narrow beam reflecting from a fixed point on a rotating
mirror, then pass through a beam splitter. The two beams then have
anticorrelated directions, and both trace back to the same
illuminated point on the mirror. The mirror, as it rotates, fills
the entire lens aperture with light over time.

The latter two methods are used
to enforce spatial correlations in quantum (\cite{16}) and
classical (\cite{23}) versions, respectively, of
correlated-photon or "ghost" imaging. In all three
methods, the light would be focused at a point satisfying ${1\over
{z_1}}+{1\over {z_2}}={1\over f}$. (In method 1, we would have $z_1=\infty$ and
$z_2=f$.)
In all of these setups we can think of the arrangement as
providing an {\it effective} or {\it virtual pinhole}, even though
there is no pinhole physically present. In each case, we may
proceed formally in the same manner that we would if there was an
actual source pinhole.

\begin{figure}
\centering
\includegraphics[width=7cm]{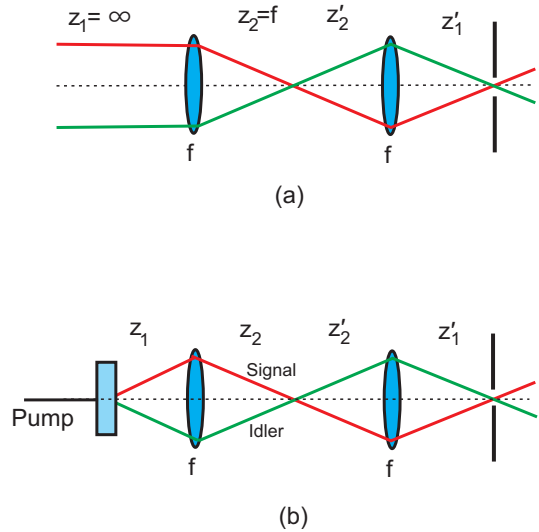}
\caption{\textit{ (Color online) Two methods for making a confocal
microscope without a source pinhole.}}
\end{figure}

There is a problem with these approaches, though.
Although correlation is maintained up to the location of the first
lens, diffractive effects at the lens then destroy a portion of
the correlation before the sample is reached. Taking into account
the possibility of aberration in the lens as well, the benefits
deriving from these methods seem to be limited.

However a completely different class of solution seems more promising: we can take
the light to be initially uncorrelated when it enters the
microscope, but then select out pairs of
photons which happen to be at the same spatial distance from the
axis, thus reducing the likelihood of detecting pairs which are at
different transverse distances. This effectively enforces
correlation at the location of the sample, where it is needed.

In order to enforce the correlation, we need to tag the photons in
some manner according to where they happen to intersect the
sample, selecting the value of some property of the photon in a
manner that depends on position in the sample plane. One way to do
this would be to construct filters with narrow, position-dependent
pass bands; then, if broadband light is used for illumination, the
light leaving the sample will be segregated, with photons of
different frequency leaving different points of the sample. Thus,
the frequency of the detected light provides an indicator of
position and can be used to enforce spatial correlation.

This frequency-tagging method will be discussed in more detail elsewhere.
In the current paper, we instead discuss in detail an
apparatus in which we tag our photons, not by frequency, but by phase. The two beams are
then combined at a beam splitter, converting the phase-tagging into correlation.
Effectively, we have two
confocal microscopes in the two branches of a Mach-Zehnder interferometer;
the interferometer serves to
compare the photon phases and to suppress pairs that differ by a large
phase. A pair of detectors and a coincidence counter then count
the number of pairs that survive this comparison.

Note that interferometry is not an essential element of the basic
correlation confocal scheme, but only of the particular embodiment of
it that we discuss in the following sections; it appears only because we chose to
tag the photons by phase. If we use frequency-tagging
instead, no interferometry would be required.
The frequency scheme could be implemented, for example, by
spatially separating different frequencies at the output with a
diffraction grating; then a coincidence count would occur when the
pixels at the same relative position are simultaneously triggered
in two CCD cameras.

In the next section, we introduce correlations into a
confocal microscope via matching of phases and then examine
the use of the resulting correlation confocal microscope to scan
over a sample.

\section{The Correlation Confocal Microscope.}

Here we introduce the correlation confocal microscope. To make the
operating principle clearer, we initially work out the formulas
for an "unfolded" version of the microscope in which the apparatus
has two branches and two identical objects, one in each arm; later
we show how to reduce this apparatus to the more useful version
with a single arm and single sample. The basic setup is shown in
figure 3.  The locations in the transverse planes at the
first lens, second lens, and image plane are represented respectively
by ${\mathbf x^\prime}$, ${\mathbf x^{\prime\prime}}$, and
${\mathbf x}$.  Note that ${\mathbf x^\prime}$ and ${\mathbf q}$
are related by ${{\mathbf q}\over k}={{{\mathbf x^\prime}}\over
{z_2}}$. Subscripts $1$ and $2$ on ${\mathbf q}$, ${\mathbf
x^\prime}$, or any variable other than $z$ will denote which
branch (upper or lower) of the apparatus is being referenced.
$t_1$ and $t_2$ represent the effect of the samples $S_1$ or $S_2$
in the two arms. In the object plane, ${\mathbf y^\prime}$ will
denote an arbitrary point in the plane, and ${\mathbf y}$ will be
the point actually being viewed in the object; in other words,
$-{\mathbf y}$ is the displacement vector of the object during the
scan. (The minus sign appears in our convention in order to keep
the image from being inverted; alternatively, $+{\mathbf y}$ could
be thought of the displacement of the {\it microscope}, with the
sample is held fixed.)

The arguments
${\mathbf y_1}$ and ${\mathbf y_2}$ of the sample functions
$t_1$ and $t_2$ will be partially linked by the spatial correlation that
we will impose; we will find that this leads to an improvement of
the lateral or transverse resolution.

\begin{figure}
\centering
\includegraphics[width=7cm]{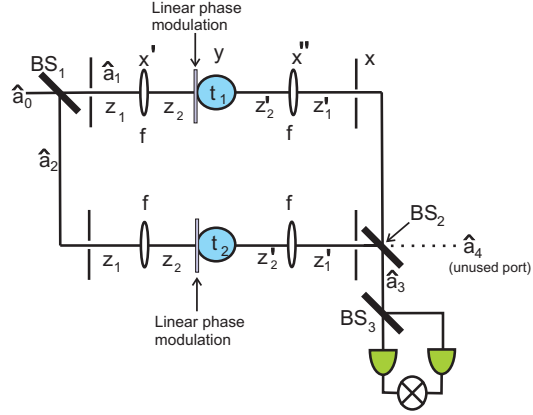}
\caption{\textit{(Color online) Schematic diagram for correlation
confocal microscope with two branches, comprised of two confocal
microscopes whose output is combined in a correlated manner.}}
\end{figure}

In front of the sample in each branch, introduce a linear position-dependent
phase shift \begin{equation}e^{i\phi_1({\mathbf y})}=e^{i (c-{\mathbf
y}\cdot {\mathbf b} ) }, \qquad e^{i\phi_2({\mathbf y})}=e^{-i
(c-{\mathbf y}\cdot {\mathbf b } ) },\label{phase}\end{equation}
for some constants $c$ and ${\mathbf b} $. We assume that ${\mathbf b}$ is a
radial vector pointing out from the axis, so that the
$\phi_i({\mathbf y})$ depend only on the magnitude $y=|{\mathbf
y}|$. The necessary phase shifts may be produced by a graded index
material; for a material $.1\; mm$ to $1\; mm$ thick, the gradient
of index versus radial distance required is of the order of
${{\Delta n}\over {\Delta r}}\sim 10^{-4}-10^{-3}\; mm^{-1}$,
which is well within the range of what is currently
technologically feasible. Alternatively, the refractive index
gradient could be achieved by inserting conically-shaped optical
wedges in front of each sample.

We now go through the operation of this apparatus. Uncorrelated
photons are input at the left. Suppose for simplicity that they
are all in the same state, represented by the creation operator
$\hat a_0^\dagger$. The first beam splitter ($BS_1$) splits the
beam between the upper arm (1) or lower arm (2) with equal
probabilities. The creation operators in these branches are
related to the initial state creation operator by
\begin{equation}\hat a_0^\dagger = {1\over \sqrt{2}}\left(
\hat a_1^\dagger-i\hat a_2^\dagger\right).\label{a0}\end{equation}

Given photons in each of the two branches (1 and 2), the creation
operators are multiplied by the impulse response functions for
passage through the confocal microscope in that branch. The creation operators for the
upper ($j=1$) and lower ($j=2$) branch just before the second beam splitter are then given
\begin{equation}\hat b_j^\dagger = \hat a_j^\dagger \int \tilde p^2\left({{k{\mathbf y_j}}\over {z_2}}\right)
t_j({\mathbf y_j}+{\mathbf y})
e^{i\phi_1({\mathbf y_j})}e^{{ik{\mathbf y_j}^2}\over {2z_2}}d^2y_j \label{b1}.
\end{equation} So if two photons are input at $BS_1$,
then by using equations (\ref{a0})-(\ref{b1}) we see that the
state incident on the second beam splitter is proportional to
\begin{eqnarray}& & \left( {1\over \sqrt{2}}\left(\hat
b^\dagger_1 +i\hat b^\dagger_2\right)\right)^2 |0\rangle = \;
{1\over 2}\left\{ \left[ \int \tilde p^2\left({{k{\mathbf
y_1}}\over {z_2}}\right) t_1({\mathbf y_1}
+{\mathbf y})e^{i\phi_1({\mathbf y_1})}e^{{ik{\mathbf y_1}^2}\over {2z_2}}d^2y_1\right]^2\hat a_1^{\dagger 2}\right.  \\
& & \quad \quad-\left[ \int \tilde p^2\left({{k{\mathbf y_2}}\over {z_2}}\right) t_2({\mathbf
y_2}
+{\mathbf y})e^{i\phi_2({\mathbf y_2})}e^{{ik{\mathbf y_2}^2}\over {2z_2}}d^2y_2\right]^2\hat a_2^{\dagger 2}\nonumber \\
& & \quad \quad +2i\hat a_1^\dagger \hat a_2^\dagger \int \tilde
p^2\left({{k{\mathbf y_1}}\over {z_2}}\right)\tilde
p^2\left({{k{\mathbf y_2}}\over {z_2}}\right) t_1({\mathbf
y_1}+{\mathbf y})t_1({\mathbf y_2}+{\mathbf y})\left. e^{i\left[
\phi_1({\mathbf y_1})+ \phi_2({\mathbf y_2})\right]}
e^{{ik({\mathbf y_1}^2+{\mathbf y_2}^2)}\over
{2z_2}}d^2y_1d^2y_2\right\}|0\rangle .\nonumber \end{eqnarray} We
may then substitute into this result the fact that
\begin{eqnarray}\hat a_1^\dagger &=&
{1\over \sqrt{2}}\left( \hat a_3^\dagger-i\hat a_4^\dagger\right)\\
\hat a_2^\dagger &=& {1\over \sqrt{2}}\left( \hat
a_4^\dagger-i\hat a_3^\dagger\right)  ,\end{eqnarray} where $3$
and $4$ represent respectively the used and unused output ports of
$BS_2$. Taking the inner product of the resulting expression with
the state having two photons leaving $BS_2$ through port $3$, we
have the two-photon amplitude in branch $3$:
\begin{eqnarray}A_3({\mathbf y}) &=& \int d^2y^\prime d^2 y^{\prime\prime } \tilde
p^2\left({{k{\mathbf y^\prime}}\over {z_2}}\right)
\tilde p^2 \left({{k{\mathbf y^{\prime\prime }}}\over {z_2}}\right) t({\mathbf
y^\prime}+{\mathbf y})t({\mathbf y^{\prime\prime}}+{\mathbf y})\nonumber \\
& & \times \; e^{{ik({\mathbf y^\prime}^2+{\mathbf y^{\prime
\prime}}^2)}\over {2z_2}}\left( e^{i\left[ \phi ({\mathbf
y^\prime})+\phi ({\mathbf y^{\prime \prime}})\right]}
+e^{-i\left[ \phi ({\mathbf y^\prime})+\phi ({\mathbf y^{\prime \prime}})\right]} -2e^{i\left[ \phi ({\mathbf
y^\prime})- \phi ({\mathbf y^{\prime
\prime}})\right]}\right) .\end{eqnarray} In the last expression,
we have set $\phi_1\equiv \phi$, $\phi_2\equiv -\phi$, and
assumed that the two objects or samples are identical
($t_1=t_2\equiv t$). The
amplitude may be put in the form (up to an overall multiplicative
constant):
\begin{eqnarray}A_3({\mathbf y})&=&\int d^2y^\prime d^2 y^{\prime\prime } \tilde p^2
\left({{k{\mathbf y^\prime}}\over {z_2}}\right)
\tilde p^2 \left({{k{\mathbf y^{\prime\prime }}}\over {z_2}}\right) t({\mathbf
y^\prime}+{\mathbf y})t({\mathbf y^{\prime\prime}}+{\mathbf y})
e^{{ik({\mathbf y^\prime}^2+{\mathbf y^{\prime
\prime}}^2)}\over {2z_2}}\nonumber \\
& & \times\left[ \cos \left( \phi \left({\mathbf
y^\prime}\right)+\phi \left({\mathbf
y^{\prime\prime}}\right)\right) -\cos \left( \phi \left({\mathbf
y^\prime}\right)-\phi \left({\mathbf
y^{\prime\prime}}\right)\right) +i \sin \left( \phi \left({\mathbf
y^\prime}\right)-\phi \left({\mathbf y^{\prime\prime}}\right)
\right) \right].\label{A3}\end{eqnarray} Note that the expression
in the square brackets equals $-1$ when $\phi ({\mathbf y^\prime})
=\phi ({\mathbf y^{\prime\prime}})={\pi\over 4}$ and vanishes when
$\phi ({\mathbf y^\prime}) =\phi ({\mathbf y^{\prime\prime}})=0$.
So if we arrange for $\phi({\mathbf y})$ to drop from ${\pi\over
4}$ at the axis to zero at the edge of the Airy disk, this
expression will strongly suppress values of ${\mathbf y^\prime}$
and ${\mathbf y^{\prime\prime}}$ that fall far from the axis,
near the edge of the Airy disk (i.e. at
the first zero of the Airy function). This is the key to the
resolution enhancement. (In the notation introduced earlier, this
means $c={\pi\over 4}$ and ${\mathbf b} ={\pi \over
{4R_{airy}}}\hat r.$) If the light is reflecting from the sample
(as opposed to being transmitted through it), then it will pass
through the phase modulation twice, so the modulation should be
half as large in this case.

When two photons simultaneously emerge from $BS_2$ into arm 3, the final beam splitter $BS_3$ simply
routes them (50\% of the time) to two different detectors, so that a coincidence count may be measured.
The result  is
that, up to overall constants, the coincidence rate will be:
\begin{equation} R_c({\mathbf y})= |A_3({\mathbf y})|^2 ,\end{equation}
where $A_3$ is given by equation (\ref{A3}).
Note that the
coincidence rate does not actually give the image of $|t({\mathbf
y})|^2$, as would normally be obtained from a
microscope, but rather it gives the image of $|t({\mathbf y})|^4$.
So the square root of the coincidence count must be taken before
comparison with the images from other microscopes.

As a special case, we can obtain the PSF by taking $t({\mathbf
y})$ to be a delta function. Keeping in mind the square root
mentioned above, this then gives us:
\begin{equation} PSF ({\mathbf y})= \left[ 1-\cos \left( 2\phi\left( y\right)
\right)\right]\;  \tilde p^4\left( {{k{\mathbf y}}\over
{z_2}}\right) =\sin^2\left( \phi\left( y\right) \right)\; \tilde
p^4\left( {{k{\mathbf y}}\over
{z_2}}\right),\label{corpsf}\end{equation} where $y$ is the
magnitude of ${\mathbf y}$. We see that the factor
$\sin^2\left( \phi\left( y\right) \right)$ is responsible for
the improvement of resolution compared
to the standard confocal microscope. This factor suppresses the
counting of photon pairs with values of ${\mathbf y^\prime}$ and
${\mathbf y^{\prime\prime}}$ that differ significantly from each
other; this, when combined with the factors of $\tilde p$, reduces
the contribution to the coincidence rate of points in the outer part
of the Airy disk. {\it Note that all of this comes about
because the factors involving the trigonometric functions in eq.
(\ref{A3}) provide a coupling between the integration variables
${\mathbf y^\prime}$ and ${\mathbf y^{\prime \prime}}$.} Without
this coupling, which amounts to a spatial correlation between the
photons detected from the two branches, eq. (\ref{A3}) factors
into two independent integrals, each of which looks like the
amplitude of a standard confocal microscope. {\it Thus, without
the correlation, nothing is gained that could not be obtained from
simply using a standard confocal microscope and squaring the
output intensity.}

The improvement introduced in this manner may be seen from
numerical simulations shown in figures 4-6. The dotted curves are
the transmittance of an object:
two square functions separated by a gap in fig. 4, three square functions in
fig. 5, and a more
complicated shape in fig. 6. The solid curve (blue online) gives the square
root of the coincidence rate of the correlation confocal
microscope, while the dashed curve (red online) gives the output intensity of a
standard confocal microscope. It can be seen clearly in these
figures that the correlation microscope produces output that
matches the original object significantly more closely than the
standard confocal microscope. In particular, in figures 4 and 5 the
correlation microscope clearly distinguishes the separate square objects
objects and correctly gives their relative heights, whereas the standard confocal microscope simply produces
a single blurred peak. Similarly, in fig. 6 the correlation
microscope clearly detects the presence of the "shoulder" on the
left-hand curve; the same feature is invisible to the
standard confocal microscope. Setting the sample equal to a
delta function, we obtain fig. 7, which shows the transverse
intensity PSF. It exhibits a roughly 60\% decrease in width,
compared to the standard confocal microscope.

\begin{figure}
\centering
\includegraphics[width=7cm]{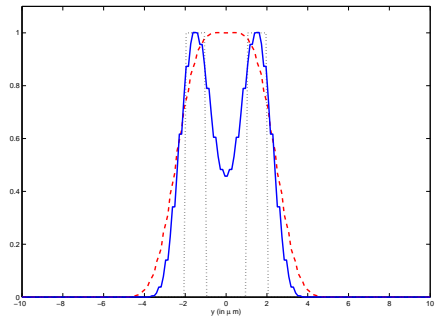}
\caption{\textit{(Color online) Comparison of images produced by
correlated (blue solid) and standard uncorrelated (red dashed)
confocal microscopes for two square objects (black dotted)
separated by a gap. The correlated version shows a substantially improved
ability to distinguish objects with small separation.}}
\end{figure}

\begin{figure}
\centering
\includegraphics[width=7cm]{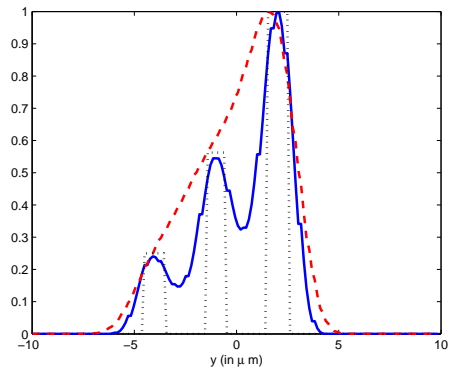}
\caption{\textit{(Color online) Comparison of images produced by
correlated (blue solid) and standard uncorrelated (red dashed)
confocal microscopes for three square objects (black dotted)
separated by gaps. }}
\end{figure}

\begin{figure}
\centering
\includegraphics[width=7cm]{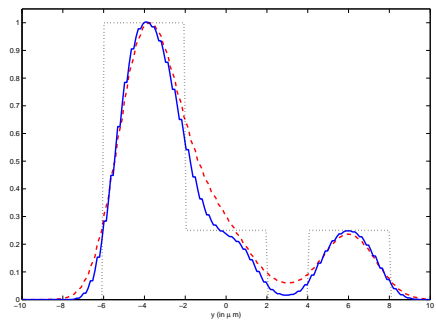}
\caption{\textit{(Color online) Comparison of images produced by
correlated (blue solid) and standard uncorrelated (red dashed)
confocal microscopes for a more complicated object (black
dotted).}}
\end{figure}

\begin{figure}
\centering
\includegraphics[width=7cm]{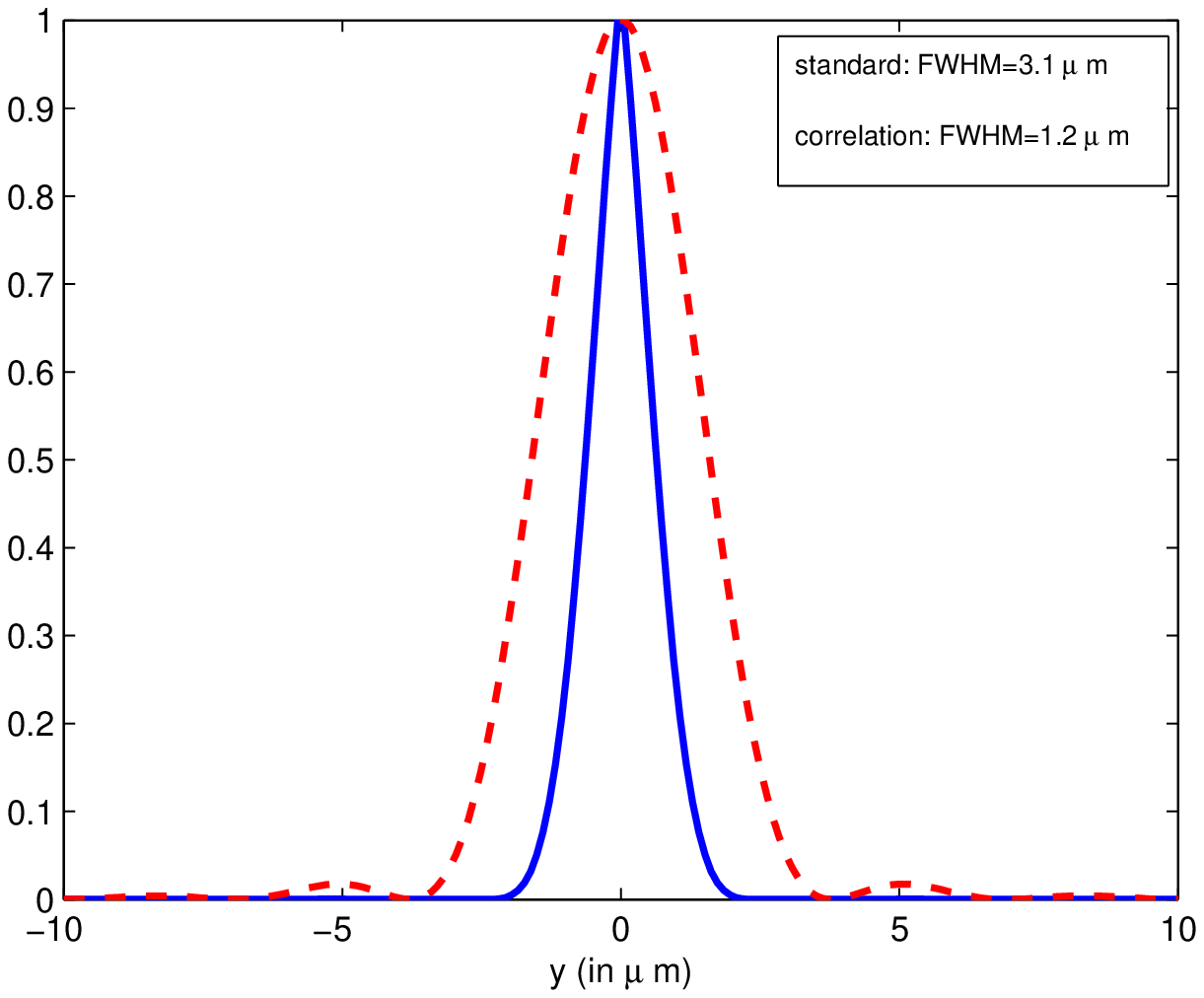}
\caption{\textit{(Color online) Comparison of transverse intensity
point spread functions for standard correlated (blue solid) and
uncorrelated (red dashed) confocal microscopes.}}
\end{figure}

Despite the substantial improvement in resolution, one problem
with this proposed microscope is also apparent: any aberrations in
the lenses or any aberrations induced by the sample will disrupt
the phase-matching, and so could reduce the benefits of this
scheme. If we chose to use the frequency-tagging scheme mentioned
earlier, instead of the phase-tagging version discussed here, a
similar degradation could be caused by frequency dispersion in the
sample. Thus, it may be advantageous to attempt finding a
variation of this setup which could incorporate dispersion
cancellation (\cite{17,18, 19}) or aberration
cancellation (\cite{20,21,22}) techniques.

\section{Reduction to One Sample.}

\begin{figure}
\centering
\includegraphics[width=7cm]{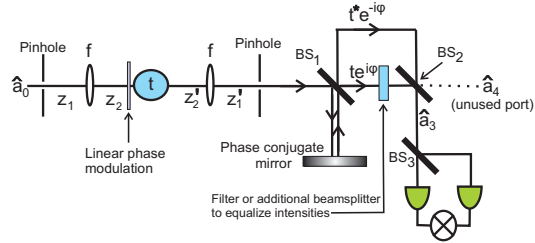}
\caption{\textit{(Color online) Correlation confocal microscope
with a single branch. }}\label{onearm}
\end{figure}

The setup in the previous section is unrealistic in the sense that
it requires two identical copies of the object or sample in order
to function. Here we go to the realistic case: we show how the
apparatus may be altered without affecting the basic principle of
operation, in order to require only a single copy of the sample.

The new setup is shown in fig. \ref{onearm}. All the light now
passes through the same sample, gaining a phase shift $e^{+i\phi
({\mathbf y})}$. But then the beam is split at
$BS_1$, after passing through the detection pinhole; half of the
beam continues onward to the second beam splitter $BS_2$
unaltered, while the other half is deflected downward to a phase
conjugating mirror. The mirror reverses the sign of the phase and
deflects the beam back to $BS_1$. The half of this beam that is
transmitted through $BS_1$ at this second encounter then
recombines with the unaltered beam at $BS_2$; from this point on,
all is as it was for the two-branch version of the previous
section, assuming $t({\mathbf y})$ is real. In the lower branch
between $BS_1$ and $BS_2$, either a high-density optical filter or
another $50/50$ beam splitter must be inserted in order to
equalize the intensity in the upper and lower branches.

If $t({\mathbf y})$ is complex,
one copy of it will also be complex conjugated at the mirror.
The factor $t({\mathbf y^\prime}+{\mathbf y})t
({\mathbf y^{\prime\prime}}+{\mathbf y})$ inside the integrals of eq.
(\ref{A3}) now becomes $t({\mathbf y^\prime}+{\mathbf y})t^\ast
({\mathbf y^{\prime\prime}}+{\mathbf y})$; since sample-induced aberrations may be viewed as extra position-dependent
phases appearing in $t$, this change, when combined with the suppression
of pairs $({\mathbf y^\prime},{\mathbf y^{\prime\prime}})$ that are far apart,
will partially cancel the sample-induced aberrations, but aberrations introduced by the lenses
are unaffected.

\section{Thick Samples and Axial Resolution.}

One of the primary advantages of the standard confocal microscope
is its sharp resolution in the axial or longitudinal direction,
allowing the ability to optically section materials, i.e. to focus
on thin slices in the interior of the material without the view
being obscured by the surrounding matter. In this section we
investigate the axial resolution of the correlation confocal
microscope.

\begin{figure}
\centering
\includegraphics[width=7cm]{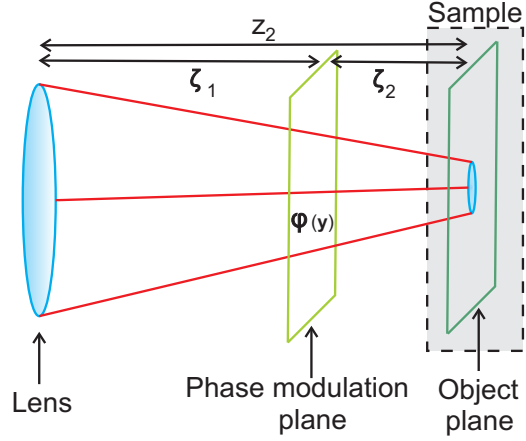}
\caption{\textit{(Color online) To view thick samples, the phase modulation plane
needs to be moved away from the object plane by distance $\zeta_2$.
}}\label{rescale} \end{figure}

Up until now, we have assumed that the plane in which the phase
shift $\phi ({\mathbf y})$ occurs is the same as the object plane;
in other words, we have taken the sample to be of negligible
thickness and the phase modulation plane to be right up against
it. But for the more realistic case of a sample of finite
thickness, with the plane to be imaged in the sample's interior, the phase
modulation obviously can not be done in the object plane. Instead,
the phase modulation plane must be moved a small distance $\zeta_2$
away from the object plane, to the exterior of the sample. The distance from the lens to this plane is
$\zeta_1=z_2-\zeta_2$ (see fig. \ref{rescale}). If
we apply a phase shift $\phi^\prime ({\mathbf x^{\prime \prime }}
)=c +{\mathbf b}^\prime \cdot{\mathbf x^{\prime\prime }}$ in this
plane, then the propagation factor from the source pinhole to the
object plane (eq. \ref{stage1}) can be shown to now be
\begin{equation} h_1(\xi =0 ,{\mathbf y})=e^{{{ik}\over
{sz_2}}y^2} e^{i(c +{\mathbf b^{\prime \prime }}\cdot {\mathbf
y})}\tilde p \left( {k\over {z_2}}\left( {\mathbf
y}-{{\zeta_2}\over k}{\mathbf b^\prime}\right)\right)
,\end{equation} where ${\mathbf b^{\prime \prime
}}={{\zeta_1}\over {z_2}}{\mathbf b^\prime }$. Note that this is
of the same form as before, except that the phase shift parameter
${\mathbf b^\prime}$ has been altered and the position of each point in the
diffraction spot has been shifted in the argument of $\tilde p $.
Since ${\mathbf b^\prime}$ is a radial vector, each point of the
Airy disk moves radially outward a distance $|\Delta y|={{b^\prime
\zeta_2}\over k}$, where $b^\prime =|{\mathbf b^\prime }|$. The
result is that the Airy disk expands in radius by $|\Delta y|$,
with a dark spot of radius $|\Delta y|$ appearing at the center.
This is clearly undesirable, so it is necessary to choose
$\zeta_2$ such that $|\Delta y|<<R_{airy}$, making the radial shift
negligible. Inserting realistic values, this means that we
$\zeta_2$ must be at most of order $10^{-4}\; m$. Thus, our
ability to view the interiors of thick samples will be limited to
depths of less than $100 \mu m$.

So, neglecting $\Delta y$ compared to $y$, we have
\begin{equation}
h_1(\xi =0 ,{\mathbf y})=e^{{{ik}\over {sz_2}}y^2} e^{i(c
+{\mathbf b^{\prime \prime }}\cdot {\mathbf y})}\tilde p \left(
{k\over {z_2}}{\mathbf y}\right) ,\label{newstage1}\end{equation}
which is identical to eq. \ref{stage1} if we choose ${\mathbf
b^{\prime\prime} } ={\mathbf b}$ (or, equivalently, ${\mathbf
b}^\prime ={{z_2}\over {\zeta_1}}{\mathbf b}$). Here, ${\mathbf
b}$ and $c$ are the original parameter values used in the previous
sections.

Suppose we view a point on the axis (${\mathbf y}=0$) at a
distance $z=z_2+\delta z$ from the lens. $z_2$ is the distance to
the confocal plane (${1\over {z_1}}+{1\over {z_2}}={1\over f}$),
so that $\delta z$ is the defocusing distance. We assume that
$|\delta z| << z_2$ and study how the axial point spread function
varies as a function of defocus. It is well known \cite{32} that
for the standard confocal microscope the axial PSF is of the form
\begin{equation} PSF_{standard}(\delta z)= \mbox{sinc}^4 \left( \kappa
\delta z\right) ,\end{equation} where $\kappa ={{ka^2}\over
{4z_2^2}}$ and $\mbox{sinc} (x)=\sin (x)/x$.

In the case of the correlation confocal microscope, each
measurement involves two photons, which may interact with the
sample at different points, so there will be two defocus
variables, $\delta z_a$ and $\delta z_b$, one for each photon.
Using eq. \ref{newstage1} in place of eq. \ref{stage1} and
allowing for defocus in each arm, a calculation similar to that of
section 4 leads to the amplitude
\begin{eqnarray} A_3({\mathbf y},\delta z_1,\delta_2) &=& \int
d^2y^\prime d^2 y^{\prime\prime }t({\mathbf y^\prime}+{\mathbf
y})t({\mathbf y^{\prime\prime}}+{\mathbf y}) e^{{ik({\mathbf
y^\prime}^2+{\mathbf y^{\prime
\prime}}^2)}\over {2z_2}} \\
& & \times E(-4\kappa \delta z_a,-{{ka}\over
{2z_2}}y^\prime)E(+4\kappa \delta z_a,-{{ka}\over {2z_2}}y^\prime)
\nonumber \\ & &\times E(-4\kappa \delta z_b,-{{ka}\over
{2z_2}}y^{\prime \prime})E(4\kappa
\delta z_b,-{{ka}\over {2z_2}}y^{\prime\prime})\nonumber \\
& & \times\left[ \cos \left( \phi \left({\mathbf
y^\prime}\right)+\phi \left({\mathbf
y^{\prime\prime}}\right)\right) -\cos \left( \phi \left({\mathbf
y^\prime}\right)-\phi \left({\mathbf
y^{\prime\prime}}\right)\right) +i \sin \left( \phi \left({\mathbf
y^\prime}\right)-\phi \left({\mathbf y^{\prime\prime}}\right)
\right) \right].\nonumber\label{newA3}\end{eqnarray} Here, we have
defined
\begin{equation} E(u,v)\equiv \left[ L(u,v)+iM(u,v)\right] =2\int_0^1
J_0(\nu \rho )e^{{i\over 2}u\rho^2} \rho d\rho ,\end{equation}
where $J_0(x)$ is the Bessel function of zeroth order. $L$ and $M$
are known as Lommel functions and can be calculated by means of
series expansions \cite{34}.

As before, the absolute square of amplitude $A_3$ gives us the
coincidence rate. If the object $t$ is taken to be constant in the
$z$ direction and a delta function in the transverse (x and y)
directions, then this coincidence rate at ${\mathbf y}=0$ will
give us the axial PSF. Up to an overall normalization constant, we
then have
\begin{eqnarray} PSF_{axial} (\delta z_a,\delta z_b) &=& \left| E(
-4\kappa \delta z_a,0) E(4\kappa \delta z_a,0) E(
-4\kappa \delta z_b,0) E(4\kappa \delta z_b,0) \right|^2 \nonumber \\
&=& \mbox{sinc}^4(\kappa \delta z_a)\mbox{sinc}^4(\kappa \delta
z_b). \end{eqnarray} The axial PSF thus factors into the
product of the axial PSFs of the two separate standard confocal
microscopes in the two arms. So, the phase modulation has no effect on
the axial resolution.

\section{Conclusions.}

In this paper we have shown that by incorporating intensity
correlations and coincidence measurements into a generalized
version of the confocal microscope, it should be possible to
achieve large improvements in resolution in the transverse
direction. This leads to a great improvement in imaging ability.

There are a number of variations of this scheme that may be
investigated, some of which may lead to further improvements. For
example, we have assumed that the phase shifts at the sample vary
linearly with radial distance; instead, a nonlinear radial
dependence could be used to provide further suppression of photon
pairs that fall far from the center of the Airy disk. Another
variation is that instead of looking for pairs of photons both
leaving $BS_2$ by the same output port (port 3), we could look for
coincidences {\it between} the two possible output ports (one
photon each in ports 3 and 4). This would simplify the apparatus;
however it can be shown that the resolution in this variation is
not as good as the one we have described in the earlier sections,
although it is still better than the standard confocal microscope.
To be specific, the improvement factor of $\left[ 1-\cos \left(
2\phi\left( y\right) \right) \right]=\sin^2\left( \phi\left(
y\right) \right)$ in eq. (\ref{corpsf}) would be replaced by $\sin
\left( 2\phi\left( y\right) \right) ,$ which decreases more slowly
away from the axis for the same choice of phase function.

Some disadvantages of the correlation confocal microscope
compared to the standard confocal
microscope should be mentioned. First, although the improved
transverse resolution leads to a smaller confocal volume, the
correlation microscope is no better than a standard confocal
microscope in fluorescence correlation spectroscopy (FCS) and
dynamic light scattering (DLS) experiments; this is because the
random phase shifts from the scattering or fluorescence events
will destroy the phase correlations. So for these experiments, the
extra complexity involved in the correlation brings no benefits.

Second, most of the photons entering the correlation confocal
microscope will be weeded out by the phase-correlation requirement
and will not contribute to the image. In this sense, the
efficiency of the device is low, and image formation will
certainly be slower than with a standard confocal microscope.
However, the substantial improvements in resolution should
certainly outweigh the loss of speed in many applications.

\section*{Acknowledgments}

This work was supported by a U. S. Army Research Office (ARO)
Multidisciplinary University Research Initiative (MURI) Grant; by
the Bernard M. Gordon Center for Subsurface Sensing and Imaging
Systems (CenSSIS), an NSF Engineering Research Center; by the
Intelligence Advanced Research Projects Activity (IARPA) and ARO
through Grant No. W911NF-07-1-0629.

The authors would like to thank Dr. Lee Goldstein and Dr. Robert Webb for some very
helpful discussions and suggestions.

\end{document}